\documentclass{article}

\usepackage[lists,markers]{endfloat}

\usepackage{graphicx}

\usepackage{natbib}

\author{%
}

\let\labak=\large \def\large{\let\bf=\sf   \let\bfseries=\relax \labak \rm\sf}
\let\Labak=\Large \def\Large{\let\bf=\sf \let\bfseries=\relax \Labak \rm\sf}
\let\LAbak=\LARGE \def\LARGE{\let\bf=\sf \let\bfseries=\relax \LAbak \rm\sf}
\let\hubak=\huge  \def\huge{\let\bf=\sf  \let\bfseries=\relax \hubak \rm\sf}
\let\Hubak=\Huge  \def\Huge{\let\bf=\sf  \let\bfseries=\relax \Hubak \rm\sf}

\advance\voffset by-0.5in
\advance\textheight by0.75in

\begin{document}
\begin{large}
\begin{center}
{\huge The dynamics of iterated transportation simulations}

\vfill

Kai Nagel,${}^{a,b,}$\footnote{Corresponding author; current
affiliation: Swiss Federal Institute of Technology (ETH) Z{\"u}rich,
Department of Computer Science, ETH Zentrum, CH-8092 Z{\"u}rich,
Switzerland; Email: kai@santafe.edu; Fax +1-810-815-1674}\ 
Marcus Rickert,${}^{a,b,}$\footnote{Current affiliation: sd\&m AG, 
Troisdorf, Germany; Email: marcus.rickert@topmail.de}\ 
Patrice M.\ Simon,${}^{a,b,}$\footnote{Current affiliation: Carlson
  Wagonlit IT, Minneapolis MN, U.S.A.; Email: psimon@Carlson.com}\
and Martin Pieck${}^{a,}$\footnote{Email: pieck@lanl.gov}

\vfill

${}^a$ Los Alamos National Laboratory, Los Alamos NM, U.S.A.\\
~\\
${}^b$ Santa Fe Institute, Santa Fe NM, U.S.A.

\vfill

This version: \today

\vfill

Earlier version presented at the TRIannual Symposium on Transportation
ANalysis (TRISTAN-III) in San Juan, Puerto Rico, 1998.

\end{center}
\end{large}

\vfill

\noindent\textbf{Abstract:}
Iterating between a router and a traffic micro-simulation is an
increasibly accepted method for doing traffic assignment.  This paper,
after pointing out that the analytical theory of simulation-based
assignment to-date is insufficient for some practical cases, presents
results of simulation studies from a real world study.  Specifically,
we look into the issues of uniqueness, variability, and robustness and
validation.  Regarding uniqueness, despite some cautionary notes from
a theoretical point of view, we find no indication of ``meta-stable''
states for the iterations.  Variability however is considerable.  By
variability we mean the variation of the simulation of a given
plan set by just changing the random seed.  We show then results from
three different micro-simulations under the same iteration scenario in
order to test for the robustness of the results under different
implementations.  We find the results encouraging, also when comparing
to reality and with a traditional assignment result.

\vfill


\vfill\eject

\section{Introduction}

Transportation-related decisions of people often depend on what
everybody else is doing.  For example, decisions about mode choice,
route choice, activity scheduling, etc., can depend on congestion,
caused by the aggregated behavior of others.  From a conceptual
viewpoint, this consistency problem causes a deadlock, since nobody
can start planning because she does not know what everybody else is
doing.


In fact, this problem is well-known not only in transportation, but in
socio-economic systems in general.  The traditional answer is to
assume that everybody has complete information and is fully
``rational'', i.e.\ that, for some given utility-function, each
individual agent picks the solution that is best for herself.  This
means that each individual agent's decision-making process now is
globally known, and so each individual agent can (in principle)
compute everybody else's decision-making process conditioned on her
own, and so she can arrive at a solution.  As a side effect, since
everybody arrives at the same solution for everybody, one can replace
the individual decision-making process by a global computation.


Since this is a fictional process, one is traditionally not interested
in the computation process itself, but just in the end result, which
is a Nash equilibrium: Nobody can be better off by unilaterally
changing strategies.  Since only the end result is of interest,
\emph{any} algorithm finding that end result (assuming it is unique)
is equally valid.  In transportation, a typical example is the user
equilibrium solution of the static assignment
problem~(e.g.~\citet{Sheffi:book}): No driver (or traffic stream) can
be better off by switching routes.  Thus, we assume that being at the
equilibrium point is behaviorally justified, and everything we do in
order to get there is just a mathematical or computational trick.


It is instructive to look at biological ecosystems for a minute.  Here
also, the behavior of everybody depends on everybody else.  For
example, an animal should not go to an area where predators catch it.
Yet, since we assume that animals are less capable than humans to
perform organized planning and reasoning, nobody ever assumed that
animals would pre-compute an optimal solution based on some utility
function.  Instead, one formulates the problem as one of
\emph{co-evolution}, where everybody's (mostly instinctive) behavior
evolves in reaction to what is going on in the environment,
constrained by the rules of genetical chemistry.

It is indeed this ``eco-system'' ($=$ agent-based) approach that more
and more groups are also taking in the simulation of socio-economic
problems.  The advantage is that one does not have to make assumptions
about properties of the system that are necessary in order to make the
mathematics work.  For example, one can just define rules on how
agents decide on switching routes, both over night and on-line, and
let the simulation run.

The disadvantage is that currently much less is known about the
dynamical properties of such systems.  The general question is how
valid such approaches are for real-world problems.  This includes the
validity of the dynamics, the uniqueness of the solution, and the
robustness of the solution towards changes in implementation.
Non-uniqueness of solutions would be annoying although it may well be
possible that this is a property of the real-world system.
Robustness, i.e.\ that similar simulation methods yield similar
results, is something we need to hope for because it will
be hard to use such simulations in practice without it.

The work behind this paper is agent-based, since it simulates all
individual entities of the traffic system, such as travelers,
vehicles, signals, etc., as separate objects, all following their own
rules and rules of interaction.  For example, there is a plan for each
individual traveler instead of origin-destination streams.  The
simulations use, however, also concepts from the traditional
equilibrium approach, notably the idea that no traveler should be able
to (significantly) improve by switching routes.  It is in general not
necessary to do this with a micro-simulation approach; we felt however
that it would be better to start out this way before moving on to more
uncertain terrain, such as truly behaviorally based decision rules.

This paper will, after a section about the problem formulation, first
review static assignment (Sec.~\ref{sec:static}) and some theoretical
results about simulation-based assignment
(Sec.~\ref{sim-based:assign}).  In this, we will argue why we consider
computational work a necessary complement to analytical progress.  We
will then proceed (Sec.~\ref{sec:context}) with a description of the
real world scenario within which our computational studies were
undertaken; we will also give a short description of the software
modules that we used.  The following sections
(\ref{uniqueness}--\ref{robustness}) then describe results, in
particular about uniqueness, variability, robustness and validation,
and about alternative comparison measures.  The paper is concluded by
a summary.

\section{Problem formulation: Dynamic traffic assignment}

The problem treated in this paper is commonly referred to as dynamic
traffic assignment, or DTA.  In general, one is given information
about the traffic network, plus a time-dependent origin-destination
(OD) matrix which represents demand.  The problem is to assign routes
to each OD stream such the the result is ``realistic'', which is
often assumed to be the same as ``in equilibrium''.

The main difference between most other simulation-based work
(e.g.~\citet{DYNAMIT,DYNASMART}) and ours is that we are interested in
an extremely disaggregated version of the problem: The transportation
network comes with information such as number of lanes, speed limits,
turn pockets, and signal phasing plans.  And the demand is given in
terms of individual trip plans, with a starting time, a starting
location on the network, and a destination.  Thus, writing this as a
time-dependent OD-matrix is not really useful: First, since each link
of the network is a potential origin or destination, one can easily
obtain a matrix with 200\,000~$\times$~200\,000~entries (TRANSIMS
Portland case-study, in preparation).  Second, our starting times come
with second-by-second resolution.  Translating this into
second-by-second OD matrixes would result in matrices which are mostly
empty, while aggregating it into longer time intervals means giving up
information.

\section{Static equilibrium assignment}
\label{sec:static}

This section contains a very short review of static equilibrium
assignment, which is the traditional approach to our problem.  The
purpose of the section is not to discuss newest developments in the
field (for a relatively recent review see, e.g.,
\citet{Patriksson:book}), but to lay the ground work to point out
certain similarities between traditional equilibrium assignment and
current implementations of simulation-based assignment.

\subsection{Deterministic User Equilibrium (UE) assignment}
\label{ue}

The traditional solution the problem of assigning traffic demand to
routes in an urban planning scenario is static deterministic
equilibrium assignment.  For our case, this would be equivalent to a
steady-state \emph{rate} of travelers for each OD pair -- anything in
our plan-set that does not correspond to a steady state rate cannot be
represented by static assignment.

Equilibrium assignment problems are usually posed in a way such that
they have a unique solution (in terms of the link flows), and
algorithms are known that come arbitrarily close to that solution.
One important assumption is that the travel time on a link is a
monotonically increasing function of the link flow -- it is this
assumption which is violated in practice since for all flow levels
below capacity there are two corresponding travel time values.  One
iterative algorithm that comes arbitrarily close to the solution is
the Frank-Wolfe algorithm.  A possible interpretation of the
Frank-Wolfe algorithm is as follows:\begin{enumerate}
  
\item Use the current set of the link travel times and
  compute fastest paths for each OD stream (also called all-or-nothing
  assignment).  If this is the first iteration, use free speed travel
  times.\label{step:one}
  
\item Find a certain optimal convex combination between the set of
  path assignments that have been computed so far and this new set of
  assignments.  In other words, for a certain fraction of the
  travelers, replace their paths by new ones.\label{step:two}
  
\item Declare this combination the current set of paths.  Calculate
  link travel times for it and start over.
  
\end{enumerate} This is repeated until some stopping criterion is met.  This
Frank-Wolfe algorithm is not the most efficient, but it is interesting
because it resembles the iterated micro-simulation technique that we
want to describe later in this text.  For more information, see
textbooks on the subject, e.g.~\citet{Sheffi:book}.

\subsection{Stochastic User Equilibrium (SUE) assignment}
\label{sue}

For stochastic assignment, one assumes that the route choice behavior
of individual travelers has a random component -- for example, because
the information is noisy, or because there is a part of the cost
function that cannot be explained by travel time, or because people's
perception is imprecise~\citep{Ben-Akiva:book}.

The standard approach to such problems is Discrete Choice
Modeling~\citep{Ben-Akiva:book}.  The outcome of this theory is that,
for a given OD pair $rs$, each alternative route $k$ is chosen with
probability $P_k$.  Faster routes are still preferred over slower
routes, but a certain fraction of the OD stream chooses the slower
routes.  Note that at this point the solution of the problem has been
made deterministic, that is, all noise is moved into the distribution
of the routes.

The solution to this is again unique under the usually assumed problem
formulation -- especially again that link travel time is a
monotonically increasing function of link flow.  An algorithm similar
to the Frank-Wolfe algorithm can be shown to be applicable.  A
possible implementation of this (for a Probit choice model) is:
\begin{itemize}
  
\item Given (in Step~\ref{step:one} of the deterministic assignment
  algorithm in Sec.~\ref{ue}) the current set of link travel times,
  compute a Monte Carlo version of an all-or-nothing assignment.
  Monte Carlo here means that, for each OD pair, we randomly disturb
  the link travel times according to the Probit distribution and
  only then calculate the fastest path.
  
\item In Step~\ref{step:two} of the deterministic assignment algorithm
  in Sec.~\ref{ue}, instead of calculating the optimal combination
  between the two sets of assignments, just use $1-1/n$ of the old set
  and $1/n$ of the new set where $n$ is the number of the iteration
  (method of successive averages). 

\end{itemize}
In other words: One uses the best current estimate of link travel
times, but then uses a noisy version of this to calculate a new
assignment; and the fraction of the old assignment to be replaced is
set to $1/n$. 

%

\section{Simulation-based assignment}
\label{sim-based:assign}

As stated above, we want to solve a problem which is highly
disaggregated and where the demand is given on a second-by-second
basis.  In addition, we assume that we are solving a real-world
problem, which means besides other things that in principle we can get 
arbitrarily realistic network information.

There is by now some agreement that such problems can be approached
with detailed micro-simulations.  That is, once we have plans --which
include starting times and exact routes-- for each traveler, we can
just feed this into a micro-simulation and extract any performance
measure such as time-dependent link travel times from the simulation.
Based on these performance measures, we can change the plans of some
or all of the travelers, re-run the micro-simulation, etc., until
some kind of relaxation criterion is met.

Let us introduce some minimal notation.  Each user $u$ chooses a route
$r_u$.  Given $N$ users, then the set of these routes is $\vec R =
(r_1, \ldots, r_N)$.  Recall that each route includes a starting time.
The resulting link costs are $\vec C = (c_1, \ldots, c_L)$, where $L$
is the number of links.  $\vec C$ depends on the time-of-day, i.e.\ 
$\vec C = \vec C(t)$.  In iterated micro-simulations, there are two
transitions:\begin{enumerate}
  
\item the simulation (also called network loading model):
\[
S: \vec R \to \vec C \ ,
\]
and 

\item the route assignment:
\[
A: \vec C \to \vec R \ .
\]

\end{enumerate}
Both mappings can be deterministic or stochastic.

Our problem can then be formulated as an iterated
map~\citep{Cascetta:89,Cascetta:Cantarella:day2day}.  One can see it
as an iterated map both in the routes or in the costs: $\vec R^n \to
\vec R^{n+1}$ or $\vec C^n \to \vec
C^{n+1}$~\citep{Bottom:thesis,Bottom:tristan}.  A fix point would be
reached if, e.g., $\vec R^{n+1} \equiv A \circ S (\vec R^n) = \vec
R^n$.  If at least one of the mappings $S$ or $A$ is stochastic, then
one cannot expect to reach a fix-point; however, one can hope to reach
a steady state density: $p(\vec R^{n+1}) = p(\vec R^n)$.

As a side remark, note that one can also formulate this as a
continuous dynamical system by making $n$ continuous; the mapping $A
\circ S$ would then be replaced by a differential equation.  Such
systems (e.g.~\citet{Friesz:etc:day2day}), although related, are
somewhat more removed from the topic of this paper since iterated
versions of dynamical systems can display vastly different dynamics
from their continuous counterparts (e.g.~\citet{Schuster:book}).

The stochastic user equilibrium case could be modeled by assuming that
we have as many ``users'' as we have routes for each OD relation $rs$,
together with a traffic stream strength $q_u^{rs}$.  The mapping $S$
is then simply the typical link cost function.  That is, link flows
are the sum of all OD streams that pass over the current link, and
link cost is a function of link flow.  The mapping $A: \vec C \to \vec
R$ would come from the particular ``re-planning'' algorithm that was
selected for the SUE problem, for example from a Monte Carlo
assignment plus the method of successive averages.  One can for
example search for a fix-point in $\vec C$, i.e.\ $\vec C^{n+1} =
C^n$.

For the more general problem of time-dependent assignment, one would
like to show similar things as one has shown for the equilibrium
assignments: for example uniqueness, and an algorithm that is
guaranteed to converge.  Indeed, it can be
shown~\citep{Cascetta:89,Cascetta:Cantarella:day2day} that under
certain circumstances the mapping $A \circ S: \vec R \to \vec R$ is
ergodic, which means that any combination of feasible routes will
eventually be used by the iterations.  ``Feasible routes'' is a set of
routes that brings the traveler from her starting location to her
destination; in general, it is a finite set since one assumes that
routes are loopless.  This means that every combination of routes has
a time-invariant probability $p(\vec R)$ to be used, and since the
system is ergodic, in order to obtain mean values one can replace the
phase-space average $\langle X \rangle = \sum_{\vec R} p(\vec R) \cdot
X(\vec R)$ by an iteration average $\overline X^T = 1/T
\sum_{n=i}^{i+T} X(\vec R^n)$, where $n$ and $i$ are iteration
indices.

The in our view most critical condition for this to be true is that
\begin{itemize}
  
\item \emph{either} each feasible route has a probability larger than
  zero to be selected in the following time step \citep{Cascetta:89} $(*)$
  
\item \emph{or} each set of feasible routes can be reached from any
  other set of feasible routes via a sequence of iterations
  \citep{Cascetta:Cantarella:day2day}. $(**)$

\end{itemize}

For practical applications, however, the situation is more
complicated.  We want to point out three examples of possible problems
with the analytical results.  We and others have never observed these
problems in the practice of simulation-based DTA; however, they
indicate that systematic simulations or an improvement of the theory
are necessary.  The examples come from
\citet{Palmer:broken:ergodicity}, which is an introduction to the
phenomenon of broken ergodicity, which is one of the possible problems
one might face.  The first two examples can be found in most textbooks 
on Statistical Physics. 
\begin{itemize}
  
\item First, ergodicity is actually not enough to ensure that the
  phase space density (i.e.\ the space of all possible route sets)
  becomes uniform and stationary. For example, it would be ergodic to
  sort all feasible route sets into a sequence and only allow
  transitions along the sequence.  A more stringent property called
  ``mixing'' is needed to cause any initial phase space distribution
  (which in our case is just a point: \emph{one} set of routes) to
  spread uniformly. $(*)$ will ensure mixing, $(**)$ will not.
  
\item Second, ergodicity only says something for the infinite time
  limit; it might take much longer than the age of the universe for an
  actual ergodic system to do a good job of covering the phase space.
  
\item Third, the system may show broken ergodicity.  That is, the
  system may be quasi-ergodic in a \emph{part} of the phase space,
  with very little yet non-zero probability of escaping from that
  part.  Our iterative assignment may be ``stuck'' with a particular
  type of solution for a very large number of iterations; if we do
  not run enough iterations, we will never see that there is another
  type of solution.  Sometimes, one calls these states
  ``meta-stable'', but that word makes the situation sound less
  problematic than it potentially can be.

\end{itemize}

In consequence, in this paper we want to report simulation experiments
with highly disaggregated DTAs in large realistic networks.  First, we
look into uniqueness of the simulation results.  Second, we are
interested in how ``robust'' our results are.  We want define
``robustness'' more in terms of common sense than in terms of a
mathematical formalism.  For this, we do not only want a single
iterative process to ``converge'', but we want the result to be
independent of any particular implementation.  In consequence, we run
many computational experiments, sometimes with variations of the same
code, sometimes with totally different code, in order to see if any of
our results are robust against these changes.  Part of the robustness
analysis is a validation, since we compare some results to field
measurements, where available.  Last, we will argue that there may be
better ways to compare simulations than the typical link-by-link
analysis, and show an example.  Before we do all this, however, we
need to describe our study set-up.

\section{Context}
\label{sec:context}

\subsection{Dallas/Fort Worth Case study}

The context of the work done for this paper is the so-called
Dallas--Fort Worth case study of the TRANSIMS
project~\citep{Beckman:etc:case-study}.  Most of the details relevant
for the present paper can also be found in
\citet{Nagel:Barrett:feedback}.  The purpose of the case study
was to show that a micro-simulation based approach to transportation
planning such as promoted by TRANSIMS will allow analysis that is
difficult or impossible with traditional assignment, such as measures
of effectiveness (MOE) by sub-populations (stakeholder analysis), in a
straightforward way.  In the following we want to mention the most
important details of the case study set-up; as said, more information
can be found in
\citet{Beckman:etc:case-study} and \citet{Nagel:Barrett:feedback}.

The underlying road network for the study (public transit was not
considered) was a so-called focused network, which had 14751~mostly
bi-directional links and 9864~nodes.  Out of those, 6124~links and
2292~nodes represented \emph{all} roads in a 5~miles times 5~miles
study area, whereas the network got considerably ``thinner'' with
further distance from the study area.\footnote{Note that this
  ``thinning out'' of the network was not done in any systematic way
  and is explicitely \emph{not} recommended.  It was an ad-hoc
  solution because more data was not available.}  A picture of the
focused network can be found in \citet{Nagel:Barrett:feedback}.

The TRANSIMS design specifies to use demographic data as input and
generate, via synthetic households and synthetic activities, the
transportation demand.  The Dallas/Fort Worth case study was based on
interim technology: part of the demand generation was not available
then.  For that reason, we use a standard time-dependent OD matrix as a
starting point, which is immediately broken down into individual
trips.  All trips are routed through the empty network, and only
trips that go through our smaller study area are retained.  This
base set contains approx.\ 300\,000 trips.  Note that this defines a
base set of trips for all subsequent studies presented in this paper:
All trips thrown out before can no longer influence the result of the
studies, although they may in reality.  Again, more information can be 
found in \citet{Beckman:etc:case-study} and \citet{Nagel:Barrett:feedback}.

\subsection{The micro-simulations}
\label{usims}

The above procedure does not only generate a base set of trips, but
also an initial set of routes (called \emph{initial planset}).  These
routes are then run through a micro-simulation, where each individual
route plan is executed subject to the constraints posed by the traffic
system (e.g.\ signals) and by other vehicles.  Note that this implies
that the micro-simulation is capable of executing pre-computed routes
(only very few micro-simulation had this capability when this work was
done although their number is growing), and it also implies that, in
the simulations, drivers do \emph{not} have the capability of changing
their routing on-line.\footnote{On-line re-routing is not incompatible
  with TRANSIMS technology \citep{Rickert:phd}, but it
  has not generally been implemented and studied.}
Three micro-simulations are used, all three related to the
TRANSIMS project, but with different levels of realism and different
intended usages.  We will call them TR (for TRANSIMS
micro-simulation), PA (for PAMINA), and QM (for Queue Model).
TR is the most realistic one, QM the least realistic ones of the
three.  The first two micro-simulations are based on the so-called
cellular automata technique for traffic flow (\citet{Nagel:etc:2lane}
and references therein).  The third one uses a simple queueing
model~\citep{Gawron:queue,Simon:Nagel:queue:tr}.

\textbf{The TRANSIMS micro-simulation (TR).}
TR is the ``mainstream'' TRANSIMS micro-simulation.  As said above, it
is the most realistic of the three, including elements such as number
of lanes, speed limits, signal plans, weaving and turn
pockets, lange changing both for vehicle speed optimization and for
plan following, etc.  The studies described in this paper were run on
five~coupled Sun Sparc~5 workstations which ran the micro-simulation
on the given problem as fast as real time; newer versions of this
micro-simulation also run on a SUN Enterprise~4000.  Details of TR
can be found in~\citet{Nagel:etc:flow-char} and in \citet{TRANSIMS}.

\textbf{The PAMINA micro-simulation (PA).}
The second micro-simulation, PA, uses simplified signal plans, and it
does neither include pocket lanes nor lane changing for plan
following.  Most other specifications are the same as for TR, although
differences can be caused by the different implementation.  PA is much
better optimized for high computing speed: it ran more than 20~times
faster than TR for this study, which is a combined effect of using
faster hardware (it is much easier to port to different hardware, thus
being able to take advantage of new and faster hardware much sooner),
less realism, and an implementation oriented towards computational
speed.  This micro-simulation is documented in~\citet{Rickert:phd},
\citet{Rickert:Wagner:Gawron}, and \citet{Rickert:Nagel:DFW}.


\textbf{The Queue Model micro-simulation (QM).}
The QM micro-simulation uses simple FIFO queues for the link exits.
These queues have a service rate equivalent to the link capacity.  The
main difference to other queueing models, e.g.~\citet{Simao:queueing},
is that in our model each link has a limited ``storage capacity'',
representing the number of vehicles that can sit on the link at jam
density.  This results in the capability to model queue spill-back
across intersections, a very important feature of congested traffic.

When a car enters a link at time $t_{enter}$, an expected link travel
time, $T_{free}$, is calculated using the length and the free flow
speed of the link.  The vehicle is then put into the queue, together
with a time $t_{dep1} = t_{enter} + T_{free}$ which marks the earliest
possible departure at the other end of the link.  In each time step,
the queue is checked if the first vehicle can leave according to
$t_{dep1}$, according to the capacity constraints, and according to
the storage constraints of the destination link.  The queue is served
until one of these conditions is not fulfilled.  The spirit of the
model is also similar to earlier versions of
INTEGRATION~\citep{INTEGRATION:94:Rel-1_5e}.  For further details on
QM, see~\citet{Simon:Nagel:queue:tr}.

The reason for having a model like this is that we want a
micro-simulation model that fits into the overall TRANSIMS framework
(i.e.\ runs on individual, pre-computed plans) but has much less
computational and data requirements than the other simulation models.
Indeed, QM runs on the same data as traditional assignment models,
and on a single CPU it is computationally a factor~20 faster than PA.
A parallel version is planned.

\subsection{Router}
\label{router}

Our micro-simulations run on precomputed route plans, i.e.\ on a
link-by-link list which connects the starting point with the
destination.  For our studies, we use a time-dependent Dijkstra
fastest path algorithm.  Link travel times are, during the simulation,
averaged into 15-minute bins.  These 15-minute bins give the link
costs for the Dijkstra algorithm~\citep{Jacob:etc:comp}.  The Dallas
study makes no attempt to include alternative modes of transportation,
such as walking, bicycle, or public transit.

We want to mention here that, in earlier versions, we randomly
disturbed link travel times by $\pm 30\%$ in order to spread out the
traffic.  This would be very similar to what some implementations of
Stochastic User Equilibrium (SUE) assignment (Sec.~\ref{sue}) do.  We
found, however, that this led to many undesirable paths, for example
cars leaving the freeway and re-entering at the same entry/exit.  In
general, it is rather difficult to find ``reasonable'' path
alternatives different from the optimal
path~\citep{Rilett:reasonable-paths}.  We would therefore expect that
also the standard SUE approach, when applied to large networks, would
display such unrealistic artifacts.

\subsection{Feedback iterations and re-planning}
\label{repl}

The initial planset is obviously wrong during heavy traffic because
drivers have not adjusted to the occurence of congestion.  In reality,
drivers avoid heavily congested segments if they can.  We model that
behavior by using \emph{iterative re-planning}: The micro-simulation
is run on a pre-computed planset and travel times along links are
collected.  Then, for a certain fraction, $f$, of the drivers, new
routes are computed based on these link travel times.  Technically,
each route from the old planset is read in, with probability $1-f$ it
is written unchanged into a new file, and with probability $f$ a new
route is computed given the starting time, starting location, and
destination location from the old route plus the time-dependent link
travel times provided by the last iteration of the micro-simulation.
In consequence, $f$ is the re-planning fraction.  After this, the
micro-simulation is run again on the new planset, more drivers are
re-routed, etc., until the system is ``relaxed'', i.e.\ no further
changes are observed from one iteration to the next except for
fluctuations (all three micro-simulations are stochastic).


\begin{figure}[t]
  \begin{center}
    \includegraphics[height=0.35\textheight]{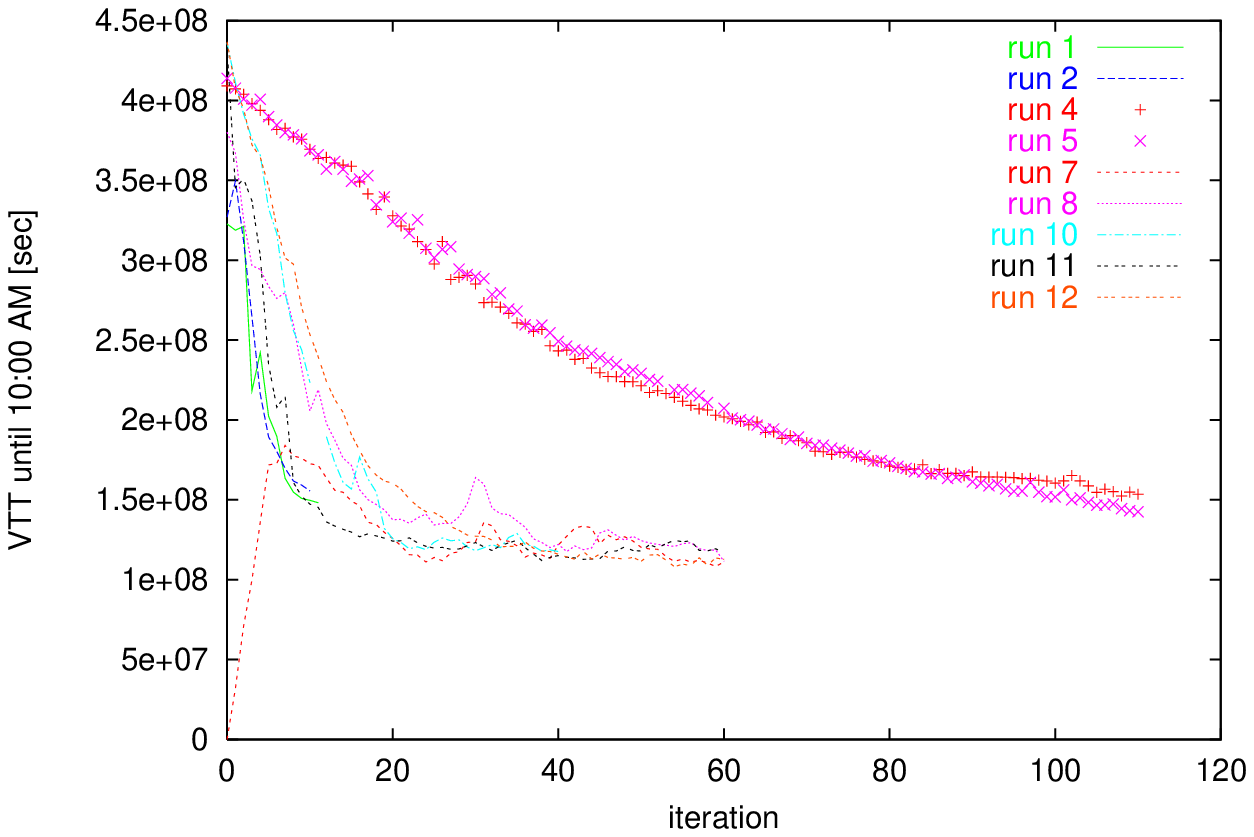}
    \caption{%
      Sum of all trip times (system-wide Vehicle Time Travelled) for
      different iteration series.  For details see
      \protect\citet{Rickert:phd}.  All iteration series relax
      towards the same VTT, which indicates (but does not prove) that
      all series releax towards the same traffic scenario.}

    \label{fig:vtt}
  \end{center}
\end{figure}

\section{Uniqueness}
\label{uniqueness}

By uniqueness we want to refer to a property that the same combination
of router and microsimulation, with the same input data, generates via
the relaxation procedure the same traffic scenario.  This still allows
for differences in the random seeds, in the way one selects travelers
for re-planning, etc.  We have run~\citep{Rickert:phd} many relaxation
experiments with different re-planning mechanisms, including an
incremental network loading over 20~iterations, a very slow iteration
series with 1\% replanning fraction, and different ways of picking the
travellers that are replanned.  We have not found any indication that
any of those relaxations ran into a traffic scenario that was
different from the other ones.

This means that in spite of the cautionary note in
Sec.~\ref{sim-based:assign} regarding broken ergodicity etc.,
practical implementations of DTA seem to be well-behaved in this
regard.  This is consistent with observations from other groups
(e.g.~\citet{Wagner:personal}) and also from experiments involving
human subjects~\citep{Mahmassani:day:to:day}.  

As a side remark, it may be worth mentioning that the best-performing
relaxation method was similar to the method of successive averages
(MSA).  What we call ``age-dependent re-planning'' \citep{Rickert:phd}
started in practice with a 30\% replanning fraction, which slowly
decreased to 5\% in the 20th iteration.  MSA by comparison uses $1/n$
as the replanning fraction, where $n$ is the iteration number.
Clearly, MSA also interpolates from high replanning fractions at the
beginning to 5\% at the 20th iteration.  However, age-dependent
re-planning also moves through the population in a systematic way,
which is more than what MSA does.  More systematic comparisons between
these methods should be tried.


\section{Variability}

Note that any given iteration $n$ corresponds to a certain set of
route plans $\vec R^n$.  For that reason, one can just re-run the
simulation of these route plans, i.e.\ just the mapping $S$.  If one
uses a different random seed, this leads to a different traffic
scenario.  \emph{These} differences (not to confuse with possible
differences caused by broken ergodicity) could be quite large in our
experiments.  An example can be found in \citet{Nagel:tgf2}, which
shows link density plots for two simulations of exactly the same route
plan sets but with different random seeds.  The microsimulation that
was used was the TRANSIMS micro-simulation, i.e.\ TR.  In one
simulation (the ``exceptional'' traffic pattern), vehicles were unable
to get off a freeway fast enough, thus blocking the freeway, thus
causing queue spillback through a significant part of the network.  In
the other simulation (the ``generic'' traffic pattern, which we also
found with many more other random seeds), this heavy queue spill-back
did not occur.

We have repeated these variability investigations with a Portland
(Oregon) scenario and a different micro-simulation -- indeed the QM
queue micro-simulation from this paper.  During those investigations,
we found that such strong variations depend, as one might expect,
heavily on the congestion level: They do not occur for low demand but
become more and more frequent when demand rises (B.\ Raney et al,
unpublished).

An interesting comparison can be made between the sources and handling
of noise in Stochastic User Equilibrium (SUE) assignment and our
method.  SUE assignment has at every iteration a ``best estimate'' of
link travel times.  A possible implementation of SUE assignment is to
take a deliberately randomized realization of those link travel times,
to re-route a fraction of the population on those, and then to take
this new combined route set and to compute the new resulting link
travel times.

Instead of deliberately randomizing our best estimate of link travel
times, we use a stochastic microsimulation which on its own generates
variability of link travel times.  The advantage of our method may be
that the noise is actually \emph{directly} generated by the traffic
system dynamics -- one should therefore assume that for example
correlations between links will be considerably more realistic as with
the parametrized noise approach of the SUE assignment.  In contrast to
SUE, however, we use the \emph{same} random realization for \emph{all}
re-planned routes.  If one route is, via a fluctuation, fast in this
realization, it will be fast for all OD pairs and thus attract a
considerable amount of new traffic.  This causes local oscillations,
which are avoided in the SUE approach.  However, remember that we
would expect unrealistic routes with some SUE implementations, see
Sec.~\ref{router}. 


\section{Robustness and Validation}
\label{robustness}

As stated above, we mean by robustness the reproducibility of results
under different implementations.  Discussion of driving rules (mostly
car following, lane changing, and gap acceptance) is a necessary part
of this, but it is not sufficient and somewhat misleading since it
does not put enough emphasis on the actual traffic outcome of the
simulation.  We propose at least two ``macroscopic'' tests:
\begin{enumerate}
\item
``Building block tests'': Test simple situations, such as traffic in a
closed loop, unprotected turn flows, etc.  See
\citet{Nagel:etc:flow-char} for a discussion of this.

\item
``Real situation tests'': Compare the results of different
micro-simulations under the same scenario.  This is the topic of this
section. 
\end{enumerate}


\begin{figure}[t]
\centerline{%
\includegraphics[height=0.4\textheight,width=\textwidth]{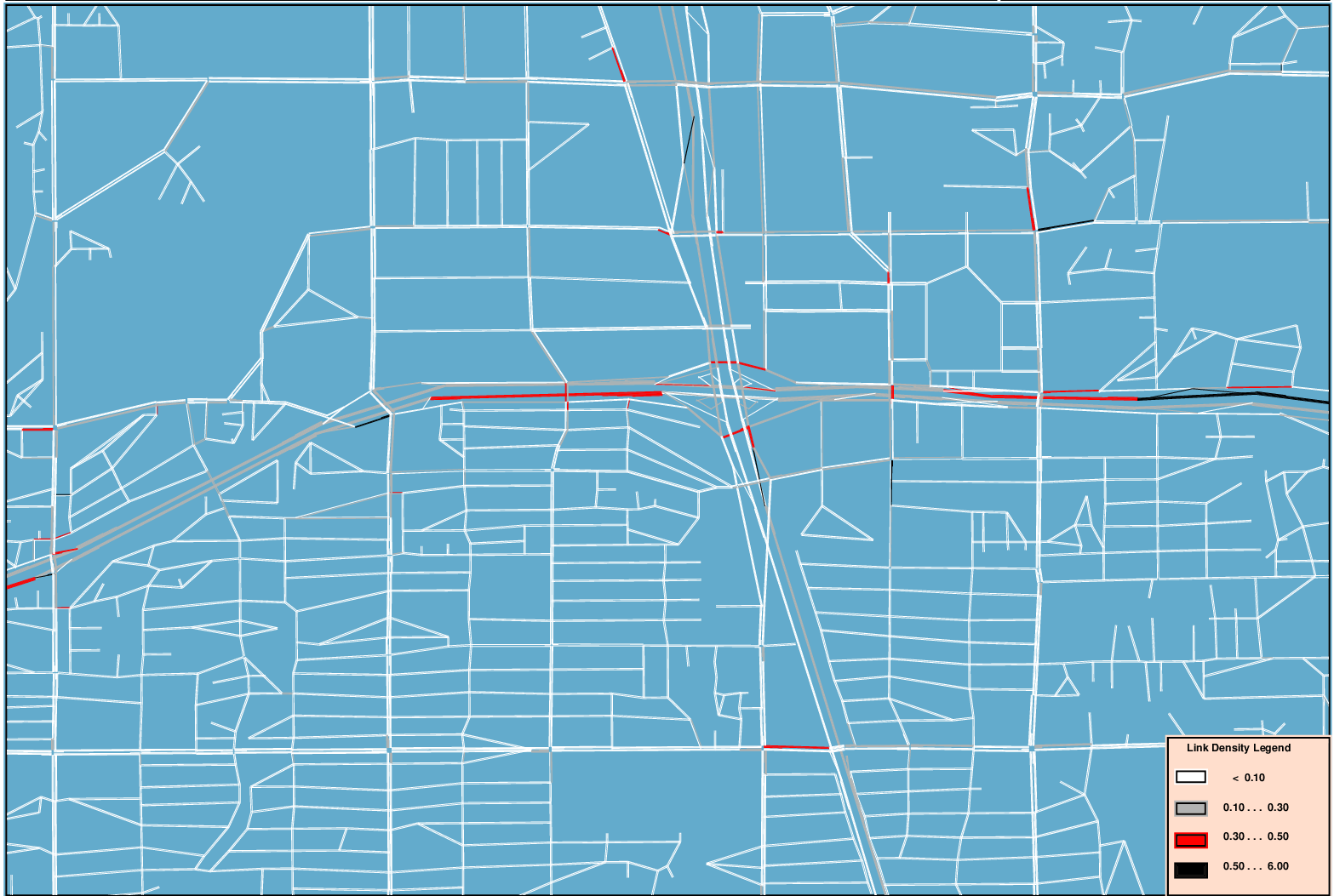}
}
\caption{TRANSIMS 14b 8:00 AM}
\label{fig_14b_8AM}
\end{figure}

\begin{figure}[t] 
\centerline{%
\includegraphics[height=0.4\textheight,width=\textwidth]{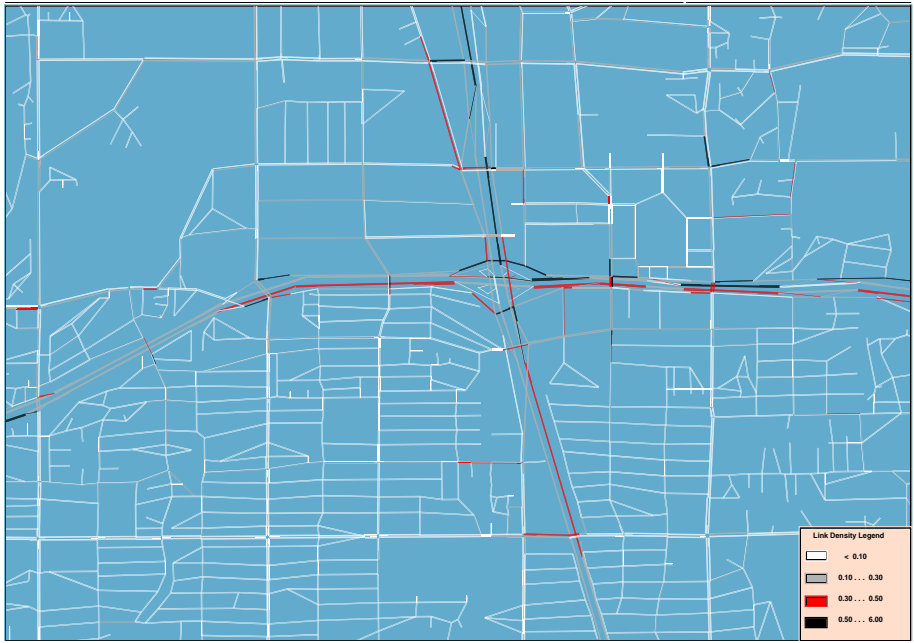}
}
\caption{PAMINA 8:00 AM}
\label{fig_pamina_8AM}
\end{figure}


\begin{figure}[t] 
\centerline{%
\includegraphics[height=0.4\textheight,width=\textwidth]{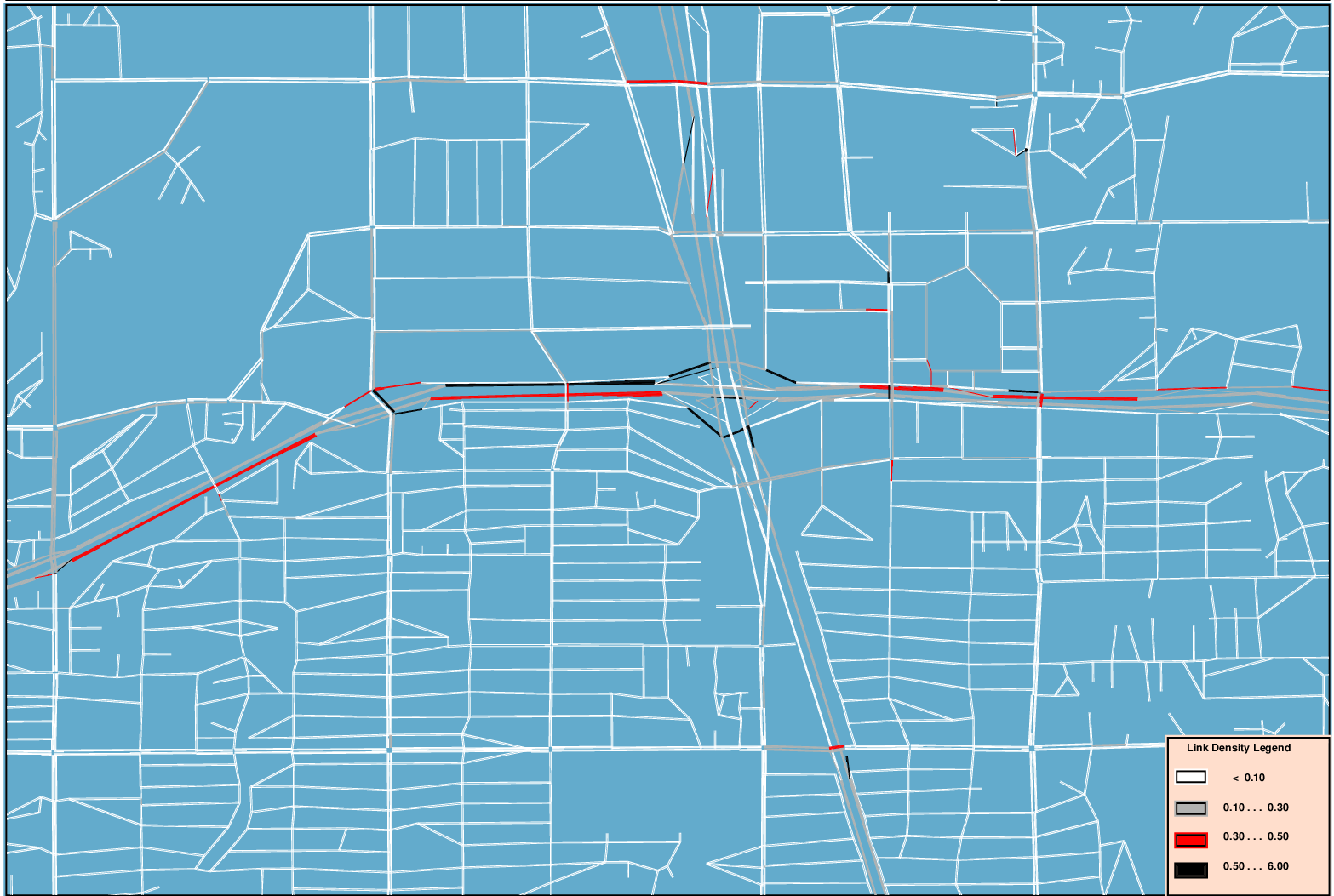}
}
\caption{QM 8:00 AM}
\label{fig_queue_8AM}
\end{figure}


\subsection{Visual comparison of link densities}

In our case, we did the same re-planning scenario, as described in
Sec.~\ref{repl}, with three different micro-simulations, as described
in Sec.~\ref{usims}.  Visual comparisons of typical relaxed traffic
patterns at 8:00am are shown in Figs.~\ref{fig_14b_8AM}
to~\ref{fig_queue_8AM}.  In our view, there is a remarkable degree of
``structural'' similarity between the plots.  This becomes
particularly clear if one compares where the simulations predict
bottlenecks, which are in general at the downstream end of congested
pieces.


\begin{figure}[t]
  \begin{center}
    \includegraphics[width=\textwidth]{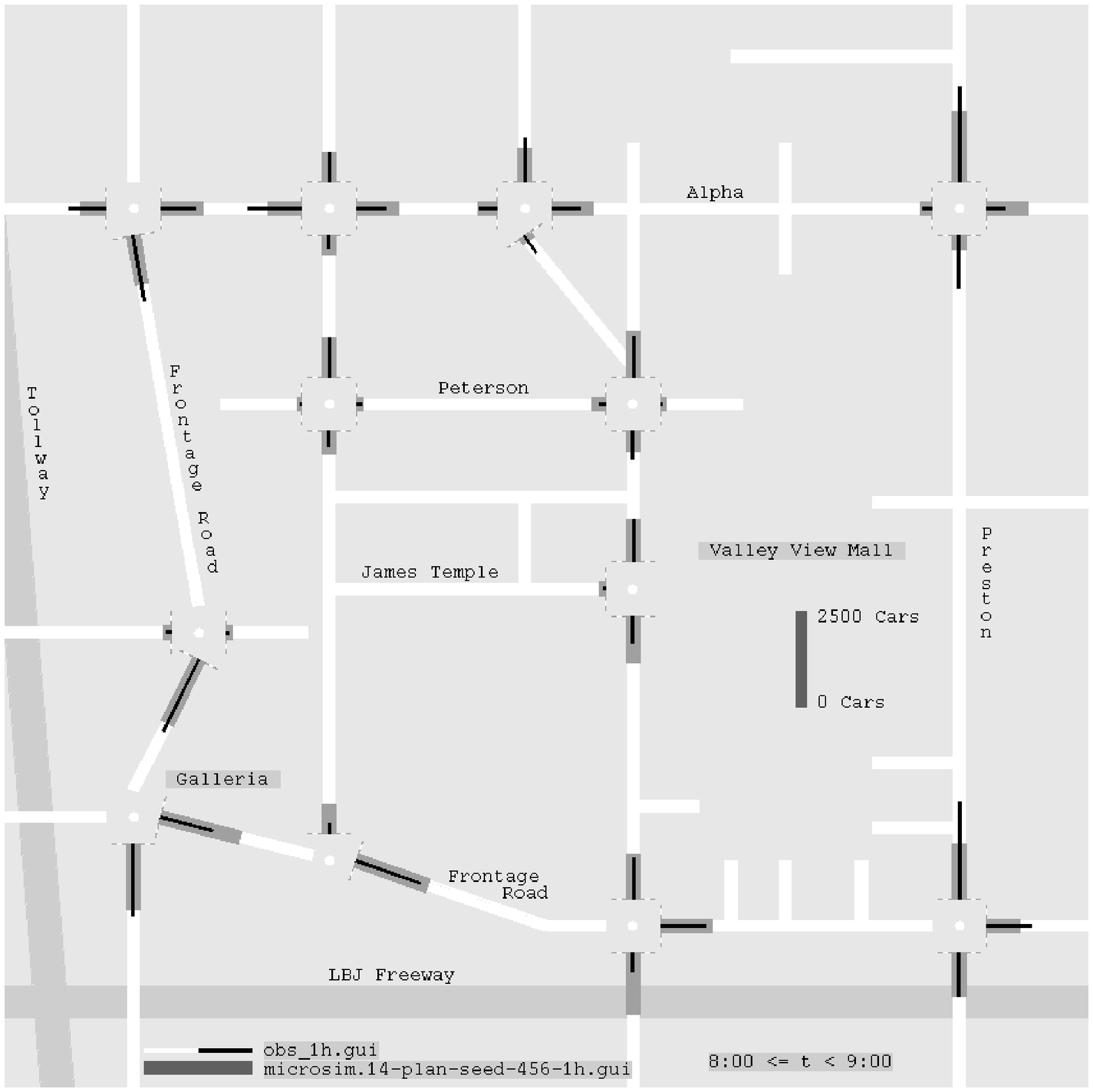}
    \caption{Approach volumes from the TRANSIMS micro-simulation series
      compared to field data.  The wide gray bars are the simulation
      results; the narrow black bars are the field data.  The freeway
      intersection in the bottom left corner of this figure is in the
      center of the study area as can be seen in
      Figs.~\protect\ref{fig_14b_8AM} to~\protect\ref{fig_pamina_8AM}.
}
    \label{fig:approach-tr}
  \end{center}
\end{figure}

\begin{figure}[t]
  \begin{center}
    \includegraphics[width=0.9\textwidth]{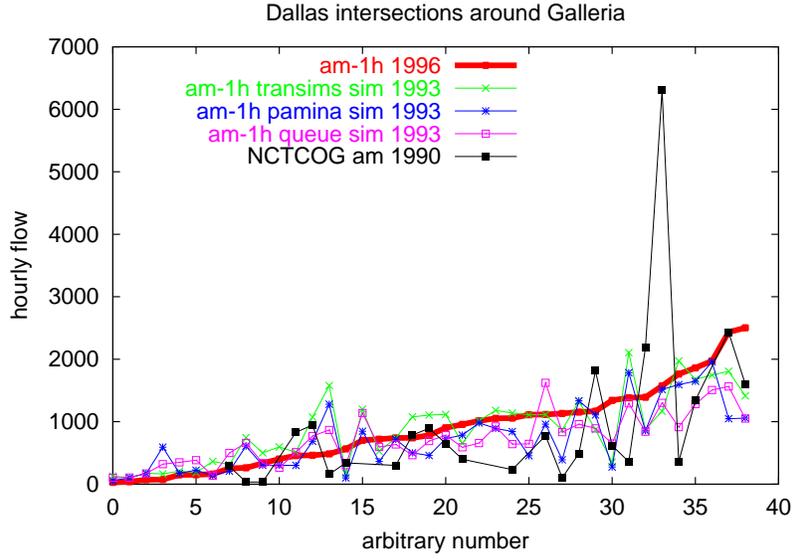}
    \caption{%
      Approach volumes for the links shown in
      Fig.~\protect\ref{fig:approach-tr}.  The links are sorted
      according to increasing flow in the field data, which is denoted
      by the wide line.  The different points in different shades of
      gray denote simulation results as denoted in the legend; the
      black line shows the NCTCOG assignment result.  Lines are
      included to guide the eye.
      }
    \label{fig:appr}
  \end{center}
\end{figure}


\begin{table}[t]
  \begin{center}
\[
\halign to\hsize{$\hfil #$\ \ && \hfil # \cr
\noalign{\hrule\smallskip\hrule\smallskip}
\hbox{Class} & $N$ & TR \hfil & PA \hfil & QM \hfil & $N_{COG}$ & NCTCOG \cr
\noalign{\smallskip\hrule\smallskip}
\le 250 : &  7 & 0.960294 &   1.217647  &   1.391176  \cr   
251 - 500 : &  7 & 1.004520 &   0.639925  &   0.577778  &   6 \hfil &  0.774523 \cr
501 - 750 : &  5 & 0.409052 &   0.354857  &   0.347362  &   3 \hfil &  0.346680 \cr
751 - 1000 : &  3 & 0.317941 &   0.252839  &   0.218395  &   3 \hfil &  0.348600 \cr
1001 - 1500 : & 11 & 0.245609 &   0.325389  &   0.307282  &   8 \hfil &  0.620177 \cr
\ge 1501 : &  6 & 0.226150 &   0.271165  &   0.370695  &   5 \hfil &  0.745831 \cr
\noalign{\smallskip\hrule\smallskip}
}
\]
    \caption{%
      Comparison between field data and simulation results.  Shown is
      the average relative error, i.e.\ $\langle
      |N_{field}-N_{sim}|/N_{field} \rangle$, where the averages are
      separate for different flow levels.  The first column shows the
      boundaries of the classes, in vehicles per hour.  The second
      column shows the number of entries per class.  The third,
      fourth, and fifth column show results for the different
      micro-simulations, TR, PA, and QM, respectively.  The sixth
      column shows again the number of entries per class, this time
      for the NCTCOG assignment; column number seven finally shows the
      results for the NCTCOG assignment.  For the TR, PA, QM, and
      NCTCOG columns, lower values are better.
      }
    \label{tab:approach}
  \end{center}
\end{table}

\subsection{Quantitative comparison of approach counts; validation}

A quantitative analysis of the above results would be useful, but is
beyond the scope of this paper.  Instead, we want to turn to link exit
volume results for a smaller number of links, which have the advantage
that comparison data from reality is available.  Note that the field
data is from 1996, whereas the demand for our simulations is from
1990.  Thus, one would expect that there is more traffic in the field
data.  Fig.~\ref{fig:approach-tr} shows a graphical comparison for the
TR micro-simulation result.  In general, it seems that we are
underestimating traffic, as we had expected.  However, we have a
tendency to overpredict traffic on low priority links.  This is
probably because our router is based on travel time only, and does not
include relevant other measures such as convenience.

Fig.~\ref{fig:appr} compares exit counts for all links that we had
field data for.  The links are sorted according to increasing flow in
the field data.  Indeed, both PA and QM are somewhat underestiming the
flows, although TR in the average does not do so.  This was also the
result of a wider comparison using other data
\citep{Beckman:etc:case-study}.

We also include results from an assignment done by the local
transportation planning authority, the North-Central Texas Council of
Governments, NCTCOG.  Unfortunately, the NCTCOG assignment data and
the field data are not really comparable since the NCTCOG assignment
was made for a different network than our simulations and the field
measurements.  In particular, the north extension of the Dallas North
Tollway had not been built, which is the freeway extending from the
interchange in the center of the study area to the north.  A result of
this is that in the NCTCOG assignment the freeway connects to a
frontage road.  This is what leads to the high assigned volume for
link~32. One should recognize that this is a physically impossible
solution since a signalized 3-lane road cannot carry 6000~vehicles per
hour.  A simulation-based method would have generated a totally
different result in the same situation.

Table~\ref{tab:approach} gives a quantitative summary of the same
data.  What is shown is the mean relative percentage deviation from
the field value, i.e.
\[
dev = {1 \over N} \sum_a { | x_a^{field} - x_a^{sim} | \over x_a^{field} } \ ,
\]
where the sum goes over all links $a$ in a class, $N$ is the number of
links in that class, and $x_a^{field}$ and $x_a^{sim}$ are the volume
counts from the field data and from the simulation, respectively.
Thus, a low value of $dev$ means a small relative difference to
reality.  According to the table, classes go from 0 to 250 vehicles
per hour, from 251 to 500 vehicles per hour, etc.  The table first
gives the class, then the number of count sites available for this
class, then the values of $dev$ for the different models.  This is
followed by data for the NCTCOG assignment, where another column of
the number of count sites is given since the NCTCOG results were only
available for a smaller number of links (that assignment was run on a
reduced network).

In spite of the difference regarding the freeway extension between the
NCTCOG result and our simulation results we believe that the
comparison between our results and the NCTCOG assignment allows the 
following two conclusions:
\begin{enumerate}
\item Our simulation-based results already at this early stage of the
  technology development yield forecast quality which is comparable to
  traditional assignment.  
\item The three micro-simulations generate results that are remarkably 
  similar in structure.  This indicates that demand generation
  research is currently at least as important as micro-simulation
  research. 
\end{enumerate}


\begin{figure}[t] 
\centerline{%
\includegraphics[height=0.4\textheight]{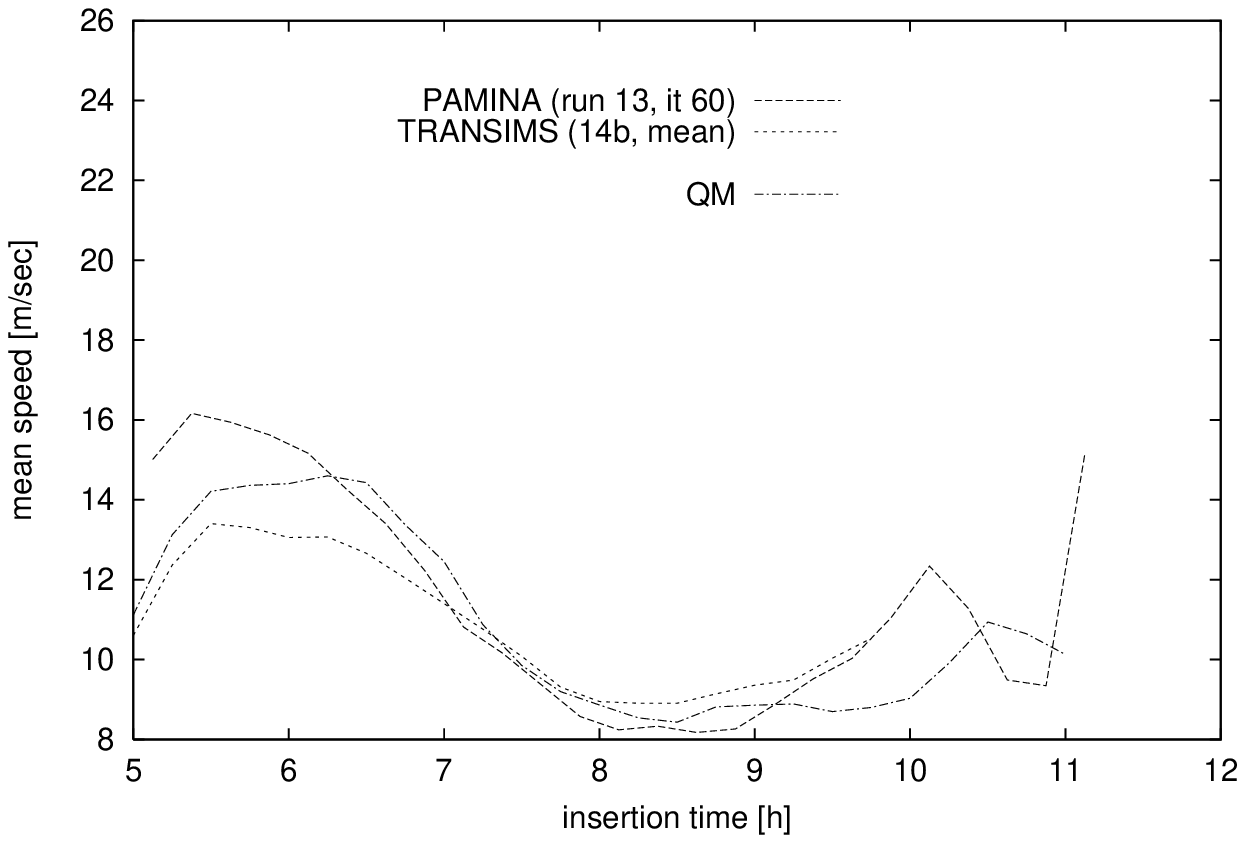}
}
\caption{``Geometrical'' mean speeds
  into the study-area, i.e.\ geometrical distance between origin and
  destination divided by trip time.  The $x$-axis shows the starting
  time for the trips.}
\label{fig_mean_speeds}
\end{figure}

\subsection{Comparisons using accessibility}

Often, transportation engineers use aggregated measures of system
performance (Measures of Effectiveness, MOEs) such as vehicle miles
traveled (VMT), the sum of all traveled distances in the system.  A
similar measure are ``geometrical'' mean speeds, which are a measure
of accessibility.  In our situation, we collected for all vehicles
with their origin outside the study area and their destination inside
the study area their geometrical distance, $d$, between origin and
destination, and their travel time, $T$.  $d/T$ is then the
``geometrical'' speed for a traveler, a measure of how fast she makes
progress towards her destination.  We see (Fig.~\ref{fig_mean_speeds})
that during the rush hour, the results of TR, PA and for QM
are practically identical.  In uncongested situations, PA predicts
faster travel than QM which predicts faster travel than TR.
This effect can be traced back to the fact
that (due to an implementation error) the maximum average speed in
TR was 75~km/h (47~MPH), in PA it was the correct design
value of 103~km/h (64~MPH), in QM it was set to the average free flow
speed which is slightly lower than PA's average speed limit.
Therefore, in uncongested situations, the predicted travel times
clearly have to be systematically different.

This indicates that aggregated measures can be considerably more
robust than more disaggregated ones.  In order to economize resources,
one should therefore pay close attention to the question at hand --
quite possibly, available models can give a robust answer for that
question even when they fail to reproduce reality on a link-by-link
basis.  This is, however, naturally also a question for further
research: Under what circumstances can we trust such aggregate
measures even when the simulation results are not close to reality on
a more disaggretaged level?

\section{Summary and discussion}

In this paper, we reported computational experiments with large scale
dynamic traffic assignments (DTA) in the context of a Dallas scenario.
An important difference of our approach to many other investigations
is that our approach is completely disaggregated, i.e.\ we treat
individual travelers from the beginning to the end.  We also used a
relatively large network, with 6124~links, where the restriction to
the network size came from data availability, not from the
capabilities of our methods.  We pointed out that although some theory
is available for DTAs, this theory needs to be used with care for
typical simulation-based DTA scenarios with a small number of
iterations.  For that reason, computational experiments remain a
necessity.

We started by looking for indications of non-uniqueness of the
solution, i.e.\ that different set-ups of the iterations could lead to
different relaxed traffic scenarios.  We did not find any indication
that this had happened in our situation.  We did, however, find
instances of very strong variability of the simulation itself, i.e.\ 
the mapping $S$ from route plans to traffic, which is stochastic.  The
reason for this is that the links are not independent; a queue which
is caused by a ``normal'' fluctuation may spill back through large
parts of the system.

We then moved to the issue of ``robustness'', by which we mean that
different implementations should yield comparable results in the same
scenarios.  In consequence, we implemented three different
micro-simulations and ran them with the same input data and the same
re-planning algorithms.  Comparisons between those results, and also
to field data and to a traditional assignment result, indicate that
(1) simulation-based assignment is already at the current stage of
research of similar quality as traditional (equilibrium) assignment,
and (2) contributions to deviations from field data come probably as
much from the demand generation as from the micro-simulations.

We concluded by arguing that a link-by-link comparison of performances
is not necessarily what one wants in order to evaluate a result.  As
an example, we showed a curve for accessibility of a certain area in
the micro-simulation as a function of time-of-day, and we pointed out
that for the critical part of the morning, which is the rush hour, the
curves for the three simulation methods are practically identic.

\section*{Acknowledgments}

KN thanks the Niels Bohr Institute in Copenhagen/Denmark for
hospitality during the time when this paper was completed.  Many
thanks to the North-Central Texas Council of Governments (NCTCOG),
especially Ken Cervenka, for preparing and providing the data.  Los
Alamos National Laboratory is operated by the University of California
for the U.S.\ Department of Energy under contract W-7405-ENG-36 (LA-UR
98-2168).

\bibliographystyle{t2} 
\bibliography{ref,kai,xref}

\begin{thebibliography}{29}
\expandafter\ifx\csname natexlab\endcsname\relax\def\natexlab#1{#1}\fi

\bibitem[{{Beckman et al}(1997)}]{Beckman:etc:case-study}
{Beckman et al}, R., 1997.
\newblock {TRANSIMS}--{R}elease 1.0 -- {T}he {D}allas-{F}ort~{W}orth case
  study.
\newblock {L}os {A}lamos {U}nclassified {R}eport ({LA-UR}) 97-4502, see
  transims.tsasa.lanl.gov.

\bibitem[{Ben-Akiva and Lerman(1985)}]{Ben-Akiva:book}
Ben-Akiva, M., Lerman, S.~R., 1985.
\newblock Discrete choice analysis.
\newblock The MIT Press, Cambridge, MA.

\bibitem[{Bottom et~al.(1998)Bottom, Ben-Akiva, Bierlaire, and
  Chabini}]{Bottom:tristan}
Bottom, J., Ben-Akiva, M., Bierlaire, M., Chabini, I., 1998.
\newblock Generation of consistent anticipatory route guidance.
\newblock In: Proceedings of TRISTAN III, vol.~2. San Juan, Puerto Rico.

\bibitem[{Bottom(in preparation)}]{Bottom:thesis}
Bottom, J.~A., in preparation.
\newblock Ph.D. thesis, Massachusetts Institute of Technology, Cambridge, MA.

\bibitem[{Cascetta(1989)}]{Cascetta:89}
Cascetta, E., 1989.
\newblock A stochastic process approach to the analysis of temporal dynamics in
  transportation networks.
\newblock Transportation Research B, 23B(1), 1--17.

\bibitem[{Cascetta and Cantarella(1991)}]{Cascetta:Cantarella:day2day}
Cascetta, E., Cantarella, C., 1991.
\newblock A day-to-day and within day dynamic stochastic assignment model.
\newblock Transportation Research A, 25A(5), 277--291.

\bibitem[{DYNAMIT(1999)}]{DYNAMIT}
DYNAMIT, 1999.
\newblock {DYNAMIT}.
\newblock Massachusetts Institute of Technology, Cambridge, Massachusetts. See
  its.mit.edu.

\bibitem[{Friesz et~al.(1994)Friesz, Bernstein, Mehta, Tobin, and
  Ganjalizadeh}]{Friesz:etc:day2day}
Friesz, T.~L., Bernstein, D., Mehta, N.~J., Tobin, R.~L., Ganjalizadeh, S.,
  1994.
\newblock Day-to-day dynamic network disequilibria and idealized traveler
  information systems.
\newblock Operations Research, 42(6), 1120--1136.

\bibitem[{Gawron(1998)}]{Gawron:queue}
Gawron, C., 1998.
\newblock An iterative algorithm to determine the dynamic user equilibrium in a
  traffic simulation model.
\newblock International Journal of Modern Physics C, 9(3), 393--407.

\bibitem[{Gawron et~al.(1997)Gawron, Rickert, and
  Wagner}]{Rickert:Wagner:Gawron}
Gawron, C., Rickert, M., Wagner, P., 1997.
\newblock Real-time simulation of the {G}erman autobahn network.
\newblock In: Proc. of the 4th Workshop on Parallel Systems and Algorithms
  (PASA `96), edited by F.~Ho{\ss}feld, E.~Maehle, E.~Mayer. World Scientific
  Publishing Co.

\bibitem[{{INTEGRATION}(1994)}]{INTEGRATION:94:Rel-1_5e}
{INTEGRATION}, 1994.
\newblock {INTEGRATION}: A model for simulating {IVHS} in integrated traffic
  networks, User's guide for model version 1.5e.
\newblock {Transportation Systems Research Group, Queens' University} and
  {M.~Van Aerde and Associates, Ltd.}

\bibitem[{Jacob et~al.(in press)Jacob, Marathe, and Nagel}]{Jacob:etc:comp}
Jacob, R.~R., Marathe, M.~V., Nagel, K., in press.
\newblock A computational study of routing algorithms for realistic
  transportation networks.
\newblock ACM Journal of Experimental Algorithms.
\newblock See www.inf.ethz.ch/\verb+~+nagel/papers.

\bibitem[{Mahmassani et~al.(1986)Mahmassani, Chang, and
  Herman}]{Mahmassani:day:to:day}
Mahmassani, H., Chang, G.-L., Herman, R., 1986.
\newblock Individual decisions and collective effects in a simulated traffic
  system.
\newblock Transportation Science, 20(4), 258.

\bibitem[{Mahmassani et~al.(1995)Mahmassani, Hu, and Jayakrishnan}]{DYNASMART}
Mahmassani, H., Hu, T., Jayakrishnan, R., 1995.
\newblock Dynamic traffic assignment and simulation for advanced network
  informatics ({DYNASMART}).
\newblock In: Urban traffic networks: Dynamic flow modeling and control, edited
  by N.~Gartner, G.~Improta. Springer, Berlin/New York.

\bibitem[{Nagel(1998)}]{Nagel:tgf2}
Nagel, K., 1998.
\newblock Experiences with iterated traffic microsimulations in {Dallas}.
\newblock In: Traffic and granular flow'97, edited by D.~Wolf,
  M.~Schreckenberg, pages 199--214. Springer, Heidelberg.

\bibitem[{Nagel and Barrett(1997)}]{Nagel:Barrett:feedback}
Nagel, K., Barrett, C., 1997.
\newblock Using microsimulation feedback for trip adaptation for realistic
  traffic in {Dallas}.
\newblock International Journal of Modern Physics C, 8(3), 505--526.

\bibitem[{Nagel et~al.(1997)Nagel, Stretz, Pieck, Leckey, Donnelly, and
  Barrett}]{Nagel:etc:flow-char}
Nagel, K., Stretz, P., Pieck, M., Leckey, S., Donnelly, R., Barrett, C.~L.,
  1997.
\newblock {TRANSIMS} traffic flow characteristics.
\newblock {L}os {A}lamos {U}nclassified {R}eport ({LA-UR}) 97-3530, see
  www.inf.ethz.ch/\verb+~+nagel/papers.
\newblock Earlier version: {T}ransportation {R}esearch {B}oard Annual Meeting
  paper 981332.

\bibitem[{Nagel et~al.(1998)Nagel, Wolf, Wagner, and Simon}]{Nagel:etc:2lane}
Nagel, K., Wolf, D., Wagner, P., Simon, P.~M., 1998.
\newblock Two-lane traffic rules for cellular automata: A systematic approach.
\newblock Physical Review E, 58(2), 1425--1437.

\bibitem[{Palmer(1989)}]{Palmer:broken:ergodicity}
Palmer, R., 1989.
\newblock Broken ergodicity.
\newblock In: Lectures in the Sciences of Complexity, edited by D.~L. Stein,
  vol.~I of Santa Fe Institute Studies in the Sciences of Complexity, pages
  275--300. Addison-Wesley.

\bibitem[{Park and Rilett(1997)}]{Rilett:reasonable-paths}
Park, D., Rilett, L.~R., 1997.
\newblock Identifying multiple and reasonable paths in transportation networks:
  {A} heuristic approach.
\newblock Transportation Research Records, 1607, 31--37.

\bibitem[{Patriksson(1994)}]{Patriksson:book}
Patriksson, M., 1994.
\newblock The Traffic Assignment Problem: Models and Methods.
\newblock Topics in Transportation. VSP, Zeist, The Netherlands.

\bibitem[{Rickert(1998)}]{Rickert:phd}
Rickert, M., 1998.
\newblock Traffic simulation on distributed memory computers.
\newblock Ph.D. thesis, University of Cologne, Germany.
\newblock See www.zpr.uni-koeln.de/\verb+~+mr/dissertation.

\bibitem[{Rickert and Nagel(1997)}]{Rickert:Nagel:DFW}
Rickert, M., Nagel, K., 1997.
\newblock Experiences with a simplified microsimulation for the {Dallas/Fort
  Worth} area.
\newblock International Journal of Modern Physics C, 8(3), 483--504.

\bibitem[{Schuster(1995)}]{Schuster:book}
Schuster, H.~G., 1995.
\newblock Deterministic Chaos: {A}n Introduction.
\newblock Wiley-VCH Verlag GmbH.

\bibitem[{Sheffi(1985)}]{Sheffi:book}
Sheffi, Y., 1985.
\newblock Urban transportation networks: Equilibrium analysis with mathematical
  programming methods.
\newblock Prentice-Hall, Englewood Cliffs, NJ, USA.

\bibitem[{Sim{\~a}o and Powell(1992)}]{Simao:queueing}
Sim{\~a}o, H., Powell, W., 1992.
\newblock Numerical methods for simulating transient, stochastic queueing
  networks.
\newblock Transportation Science, 26, 296.

\bibitem[{Simon and Nagel(1999)}]{Simon:Nagel:queue:tr}
Simon, P.~M., Nagel, K., 1999.
\newblock Simple queueing model applied to the city of {P}ortland.
\newblock International Journal of Modern Physics C, 10(5), 941--960.
\newblock Earlier version Transportation Research Board Annual Meeting paper
  99\,12\,49.

\bibitem[{TRANSIMS(since 1992)}]{TRANSIMS}
TRANSIMS, since 1992.
\newblock {TRANSIMS}, {TR}ansportation {AN}alysis and {SIM}ulation {S}ystem.
\newblock See transims.tsasa.lanl.gov.

\bibitem[{Wagner(personal communication)}]{Wagner:personal}
Wagner, P., personal communication.

\end{thebibliography}
\end{document}